\documentclass[pra,aps,twocolumn,superscriptaddress]{revtex4}
\usepackage{hyperref}
\usepackage{bm}
\usepackage{epsfig}
\usepackage{amssymb}
\usepackage{graphicx}
\usepackage{epstopdf}
\usepackage{bbold}

\usepackage{ulem}
\usepackage{amsmath}
\usepackage{dsfont}

\usepackage{color}

\begin{document}
\title{Free-space quantum signatures using 
{heterodyne} measurements}

\author{Callum Croal}
\affiliation{School of Physics and Astronomy, University of St.
Andrews, North Haugh, St. Andrews, Fife, KY16 9SS, Scotland}
\author{Christian Peuntinger}
\affiliation{Max Planck Institute for the Science of Light,
G\"unther-Scharowsky-Str. 1/Bldg. 24, Erlangen, Germany}
\affiliation{Institute of Optics, Information and Photonics,
University of Erlangen-Nuremberg, Staudtstra{\ss}e 7/B2, Erlangen,
Germany}
\affiliation{Department of Physics, University of Otago, 730 Cumberland Street, Dunedin, New Zealand}
\author{Bettina Heim}
\affiliation{Max Planck Institute for the Science of Light,
G\"unther-Scharowsky-Str. 1/Bldg. 24, Erlangen, Germany}
\affiliation{Institute of Optics, Information and Photonics,
University of Erlangen-Nuremberg, Staudtstra{\ss}e 7/B2, Erlangen,
Germany}
\author{{Imran Khan}}
\affiliation{Max Planck Institute for the Science of Light,
G\"unther-Scharowsky-Str. 1/Bldg. 24, Erlangen, Germany}
\affiliation{Institute of Optics, Information and Photonics,
University of Erlangen-Nuremberg, Staudtstra{\ss}e 7/B2, Erlangen,
Germany}
\author{Christoph Marquardt}
\affiliation{Max Planck Institute for the Science of Light,
G\"unther-Scharowsky-Str. 1/Bldg. 24, Erlangen, Germany}
\affiliation{Institute of Optics, Information and Photonics,
University of Erlangen-Nuremberg, Staudtstra{\ss}e 7/B2, Erlangen,
Germany}
\author{Gerd Leuchs}
\affiliation{Max Planck Institute for the Science of Light,
G\"unther-Scharowsky-Str. 1/Bldg. 24, Erlangen, Germany}
\affiliation{Institute of Optics, Information and Photonics,
University of Erlangen-Nuremberg, Staudtstra{\ss}e 7/B2, Erlangen,
Germany}
\author{Petros Wallden}
\affiliation{School of Informatics, University of Edinburgh, 10 Crichton Street, Edinburgh, EH8 9AB, United Kingdom}
\author{Erika Andersson}
\affiliation{SUPA, Institute of Photonics and Quantum Sciences, School of Engineering and Physical Sciences, Heriot-Watt Universiity, David Brewster Building, Edinburgh, EH14 4AS, United Kingdom}
\author{Natalia Korolkova}
\affiliation{School of Physics and Astronomy, University of St.
Andrews, North Haugh, St. Andrews, Fife, KY16 9SS, Scotland}

\date{\today}

\begin{abstract}
{Digital signatures guarantee the authorship of electronic communications. Currently used ``classical" signature schemes rely on unproven computational assumptions for security, while quantum signatures rely only on the laws of quantum mechanics. Previous quantum signature schemes have used unambiguous quantum measurements. Such measurements, however, sometimes give no result, reducing the efficiency of the protocol. Here, we instead use {heterodyne} detection, which always gives a result, although 
{ there is always some uncertainty.} We experimentally demonstrate feasibility in a real environment by distributing signature states through a noisy 1.6~km free-space channel. Our results show that continuous-variable {heterodyne} detection improves the signature rate for this type of scheme and therefore represents an interesting direction in the search for practical quantum signature schemes.}


\end{abstract}

\maketitle


Digital signatures \cite{Diffie_77} are ubiquitous in electronic communication, {used} in, for example, e-mail and digital banking. They guarantee the provenance, integrity and transferability of messages. 
Currently used classical digital signature schemes, however, rely on unproven computational assumptions~\cite{Knuth_69}, {and may become insecure} especially if quantum computers can be built \cite{Shor_97}. Quantum digital signatures (QDS)\cite{Gottesman_01, Andersson_06, Clarke_12, Dunjko_14, Collins_14, Amiri_15, Yin_15}, on the other hand, give
information-theoretic security \cite{Gottesman_01}, loosely speaking based on {the fact that non-orthogonal quantum states cannot be perfectly distinguished from each other.}

The first quantum signature schemes assumed tamper-proof, ``authenticated" 
quantum communication links. Intuitively, this could be accomplished using parameter estimation techniques similar to those used in quantum key distribution (QKD). 
How to achieve this was explicitly shown only recently~\cite{Amiri_15b, Yin_15}. {In addition, r}ecent quantum signature schemes {\cite{Dunjko_14, Collins_14}, including our protocol,} do not require {long-}term 
quantum memory. Importantly, this means that quantum signatures {can be implemented} with current technology, {essentially similar to QKD setups}. {``Classical" signature schemes with information-theoretic security also exist~\cite{chaum, hanaoka, swansonstinson}, but rely on secret shared keys, which could be accomplished using QKD. Quantum signature schemes have some advantages over such classical schemes~\cite{Amiri_15b}, but exactly what signature schemes are the most efficient remains an open problem.}

Since messages may be forwarded between recipients, a signature protocol has at least three parties, a sender Alice and two recipients Bob and Charlie. 
In QKD, the communicating parties Alice and Bob are assumed to be honest. In signature protocols, however, any of the involved parties could be dishonest. Signature schemes should be secure against forging (with high probability, only messages sent by Alice should be accepted) and against repudiation {(it is unlikely that Alice could successfully deny having sent a message that she did send)}. {Repudiation} is closely related to message transferability. {Transferability means that it is unlikely that one recipient accepts a message as genuine, but that this message then is rejected if it is forwarded to another recipient.} If there is no trusted third party, one way to settle disputes is by majority voting. For three parties, which is the case we will consider, non-repudiation and message transferability then become equivalent.  

{In principle, quantum signature schemes are based on a ``quantum one-way function" 
which maps classical information 
(a ``private key'') to non-orthogonal quantum states 
(a ``public key") \cite{Gottesman_01}. In the simplest case, Alice wants to be able to later on send a one-bit message ``0" or ``1". For longer messages, the scheme could be suitably iterated. 
Generically, signature schemes have a {\it distribution stage}, where the scheme is set up, and a {\it messaging stage}, when messages are sent and received. The distribution stage could be compared to leaving a sample of a handwritten signature e.g. when first opening a bank account. The messaging stage typically takes place much later. In our quantum signature scheme, the messaging stage is entirely ``classical".}

{In the distribution stage, Alice selects sequences 
of quantum states, one sequence for each possible future message ``0" and ``1".  The states in the sequences are selected from some set of non-orthogonal quantum states. The classical information about what states Alice has selected forms her ``private keys" for the possible messages ``0" or ``1". The quantum state sequences are the corresponding ``public keys". Alice then sends copies of the ``public key" sequences 
to Bob and Charlie, who measure the states they receive. Since it is impossible to perfectly discriminate non-orthogonal quantum states, Bob and Charlie, or any other party, can never obtain full information about Alice's ``private keys''.} 

{Later on, in the messaging stage, when Alice wants to send a message to Bob or Charlie, she sends the message together with the corresponding ``private key". 
The recipient of a message checks that the appended private key sufficiently well matches the measurement results he obtained in the distribution stage} { for the respective message}{. In a real implementation, there will be mismatches even for a private key sent by an honest Alice. However, if imperfections are not too high, then anyone other than Alice would cause a higher level of mismatches than Alice}
{. This guarantees security against message forging.} 

{Similarly, to forward a message, a recipient forwards the message together with its private key, received from Alice, and the new recipient checks for mismatches with his measurement record.
Related to this, Bob and Charlie also need to ensure that Alice cannot cheat, which would mean that she could make them disagree about the validity of a message. They achieve this by some kind of symmetrization procedure, done in the distribution stage~\cite{Gottesman_01, Andersson_06, Wallden_15}. In our protocol, as in~\cite{Wallden_15}, Bob and Charlie randomly forward half of their obtained measurement results to each other using a classical communication channel, secret from Alice. This channel could be realized using standard quantum key distribution. To ensure that Alice is unlikely to make Bob and Charlie disagree about the validity of a signature, the threshold for accepting a message directly from Alice should be stricter than for accepting a forwarded message. For more details see \cite{supplementary}.}

In this paper, we implement a quantum signature scheme using continuous variable (CV) {heterodyne} quantum measurements. Previous quantum signature schemes~\cite{Clarke_12, Collins_14, Donaldson_15}  have instead used unambiguous quantum measurements. We demonstrate that our scheme is viable in a noisy environment using a free-space urban optical communication link. Finally, we show that, even when experimental imperfections are taken into account, this scheme outperforms {a recent scheme that uses unambiguous state elimination measurements~\cite{Donaldson_15}.}


\begin{figure}[tb]
\includegraphics[width=8cm]{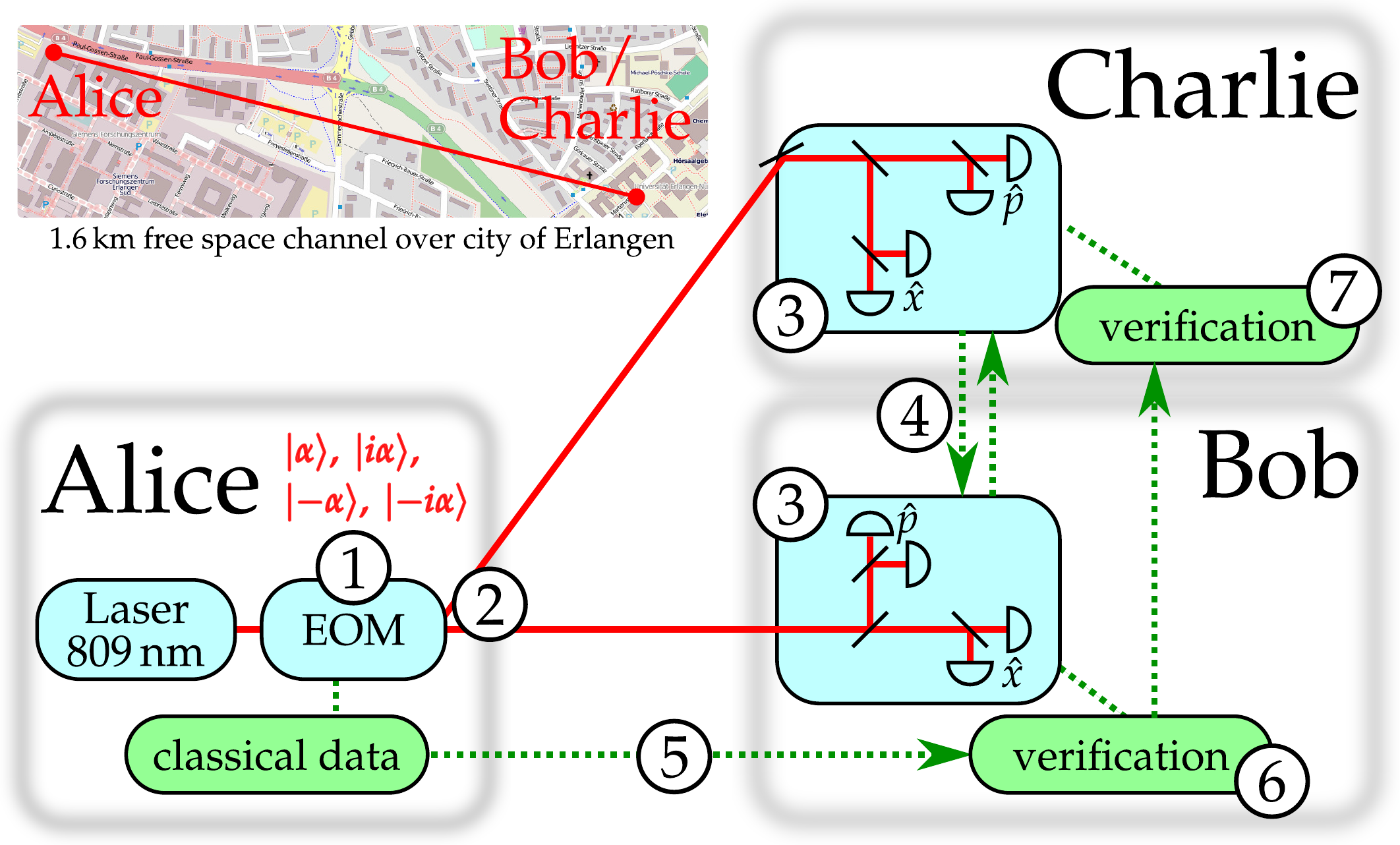}
\caption{Depiction of the scheme. The numbered parts relate to the corresponding stages in the main text. Green dashed lines indicate classical communication. Red lines indicate communication with quantum states. }\label{schematic}
\end{figure}

Our QDS scheme is represented in Fig. \ref{schematic}, with the protocol described below.
The stages  in the text correspond to the respective numbers in the figure. We use a discrete set of CV states, four phase-encoded coherent states $|\alpha\rangle$, $|i\alpha\rangle$, $|-\alpha\rangle$, $|-i\alpha\rangle$, and {heterodyne} CV measurements~\cite{Leonhardt_10}. These same states {were also} used in previous QDS schemes \cite{Clarke_12, Collins_14, Donaldson_15} and are similar to those used in some types of CV QKD \cite{Leverrier_09,Lorenz_04}.  
In \cite{Clarke_12, Collins_14, Donaldson_15}, however, recipients made ``discrete" quantum measurements with error-free (unambiguous) results, at the expense of sometimes obtaining no result. Here we instead perform {heterodyne} measurements, which always give a result, at the expense of increased errors in the results. In many cases, unambiguous results are required for a protocol to perform efficiently \cite{Bennett_92, Bergou_03}. Surprisingly, we find that for this particular QDS protocol, {heterodyne} measurements provide an advantage.

\noindent {\it Distribution stage: 1-4}

\noindent 1. For each possible future one-bit message $k=0,1$, Alice generates two identical copies of sequences of phase-encoded coherent states, $QuantSig_k=\otimes_{l=1}^L$
{$|\psi_l^k\rangle\langle\psi_l^k|$, where $|\psi_l^k\rangle$} is a randomly chosen phase-encoded coherent state, $|\psi_l^k\rangle=|\alpha e^{i\phi_l^k}\rangle$, $\phi_l^k\in \{0$, $\pi/2$, $\pi$, $3\pi/2\}$, and $L$ is a suitably chosen integer. The state $QuantSig_k$ is called the quantum signature, and the sequence of phases $PrivKey_k=(\phi_1^k,...\phi_L^k)$ is called the private key.

\noindent 2. Alice sends one copy of $QuantSig_k$ to Bob and one to Charlie, for each possible message $k=0$ and $k=1$.

\noindent 3. Bob (Charlie) measures the states received from Alice by performing a {heterodyne} detection \cite{Leonhardt_10,Leonhardt_95} of the $\hat{x}$- and $\hat{p}$-quadrature
. He records the result of the measurement and the associated position {in the sequence} $l$. 
For each quadrature, the sign of the measured result determines which state is eliminated. For example if a positive result is measured, then the state $|-\alpha\rangle$ or $|-i\alpha\rangle$ is eliminated, depending on the measured quadrature. In this way, Bob (Charlie) eliminates two states, one for each quadrature, for each signature element. 

\noindent 4. {Symmetrization:} Bob (Charlie), for each element $l$ of $QuantSig_k$, randomly chooses with equal probability to either forward the measurement results and position to Charlie (Bob) or not, secret from Alice, who should not learn the positions of the forwarded results. The resulting sequences of measurement outcomes, after the forwarding procedure, form Bob's and Charlie's ``eliminated signatures". Bob (Charlie) keeps the results obtained directly from Alice, and the results forwarded to him by Charlie (Bob) separate. Therefore, he has an eliminated signature in two parts, each of length $L/2$.

{Hom}odyne measurements will, even in the ideal case, sometimes eliminate the sent state.
If everybody follows the protocol, the probability for this depends on the {overlap} of the coherent states, 
and would  be equal to $
\frac{1}{2}\mbox{erfc}\left({\alpha}/\sqrt{2}\right)$ in the ideal case with no loss or experimental imperfections, where erfc($x$) is the complementary error function. For $\alpha=0$, this probability equals one half, and quickly approaches zero as $\alpha$ increases. Due to the unavoidable errors, this measurement protocol is an example of ``ambiguous state elimination".
Since measurements are performed immediately on receipt of the states, no quantum memory is required, just as in \cite{Collins_14,Dunjko_14}.

\noindent {\it Messaging stage: 5-7}

\noindent 5. To send a signed one-bit message $m$, Alice sends $(m, PrivKey_m)$ to Bob.

\noindent 6. Bob checks whether $(m,PrivKey_m)$ matches both parts of his stored eliminated signature by counting how many elements of Alice's private key were eliminated during the distribution stage. If there are fewer than $s_aL/2$ mismatches in each of the two parts of his eliminated signature, where $s_a$ is the authentication threshold, Bob accepts the message. 

\noindent 7. {If Bob wishes to forward a message, he forwards the message and its corresponding private key.} Charlie tests for mismatches in the same way as Bob, but with a higher verification threshold $s_v$, to protect against repudiation. Charlie accepts the message if there are fewer than $s_vL/2$ mismatches in each of the two parts of his eliminated signature, 
with $p_{err}< s_a<s_v<\frac{1}{2}$.


\begin{figure}[tb]
\includegraphics[width=8cm]{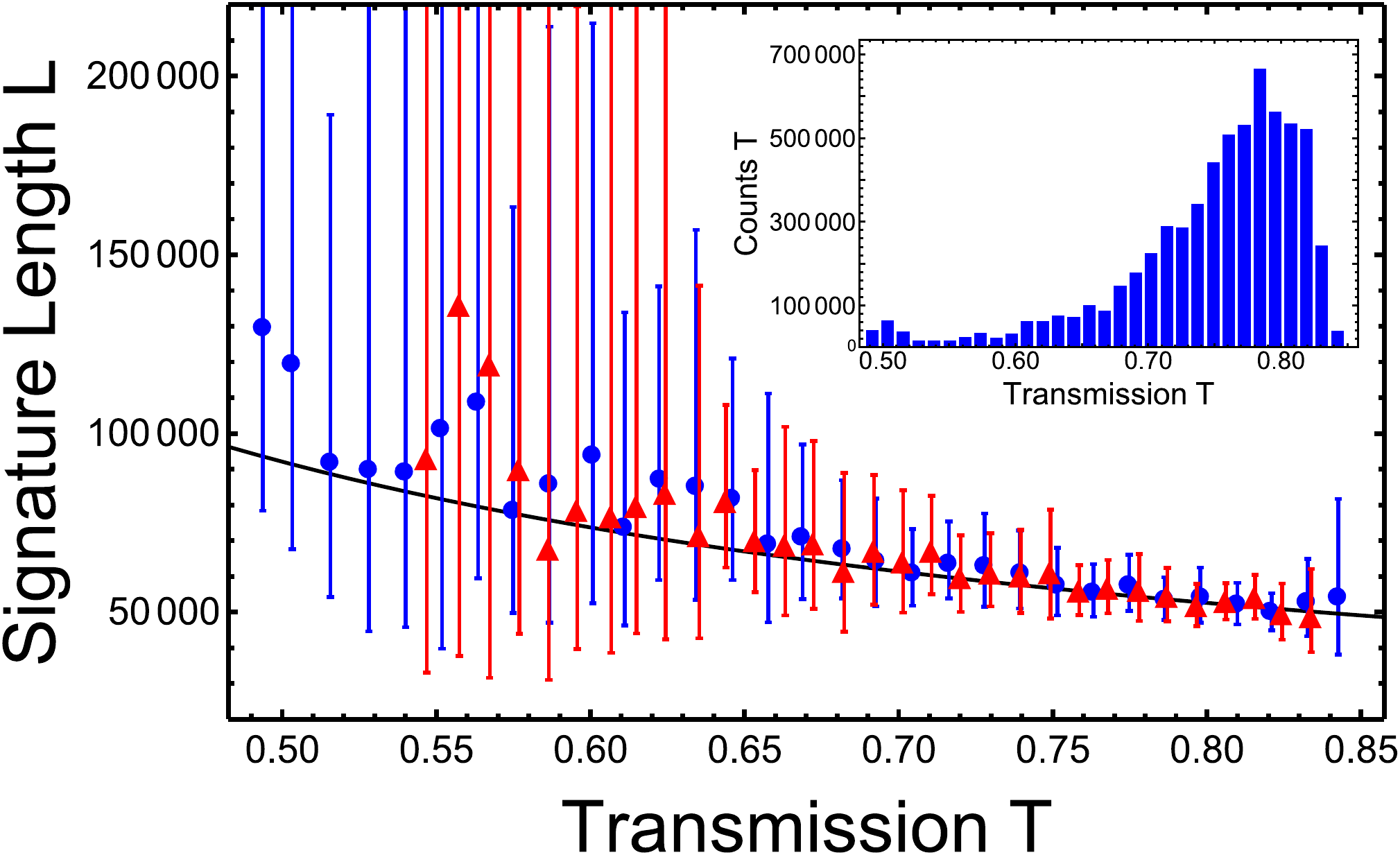}
\caption{Signature length for $\alpha=0.48$. Blue curve: theoretical model. Blue dots/bars: results from the data attributed to Bob. Red triangles/bars: results from the data attributed to Charlie. The error bars calculated are {derived by investigating the standard deviation of ten subsets of the
entire dataset}. The errors naturally increase with decreasing transmission since $g$ from Eq. \eqref{Pfailure} decreases. 
{In addition, }less data was available at lower transmission values {(see histogram of signals received by Bob per transmission sub-channel as inset)}. The data used for each point comes from a small range of transmissions, but horizontal error bars are omitted for clarity.}\label{fig1}
\end{figure}

{In essence, the security of this scheme comes from two sources. 
First, 
it is impossible for a forger to perfectly determine the {private key, since the used quantum states are non-orthogonal}. If noise is sufficiently low, the distributor Alice has an advantage over any other party. Second, the forwarding of 
measurement results 
ensures that, from Alice's point of view, Bob's and Charlie's measurement records follow the same statistics. 
This means that if Charlie uses a higher verification threshold $s_v$ than Bob's authentication threshold $s_a$, then Alice's probability to  repudiate can be made arbitrarily small by choosing the signature length $L$ large enough.
An upper bound on the repudiation probability is calculated using the Hoeffding inequality \cite{Hoeffding_63}  in the supplemental material \cite{supplementary}.}

Security against collective attacks follows from the fact that different signature states are completely uncorrelated, meaning that the optimal collective attack is an individual attack on each signature element \cite{Collins_14}. Security against coherent attacks is left for future work, noting that due to the forwarding of measurement results amongst other things \cite{Wallden_14}, methods from the security of QKD cannot be directly carried over. {Security against coherent attacks has nevertheless been analysed for a related quantum signature protocol~\cite{Wallden_15,Amiri_15b}.} We also assume that there are authenticated quantum channels between Alice, Bob and Charlie. Some kind of parameter estimation procedure should be used to replace this assumption, analogous to~\cite{Amiri_15b, Yin_15}.

To successfully forge, Bob must guess a sequence of states that meets Charlie's verification threshold. 
For individual and collective forging, the optimal forging attack is to perform a minimum-cost measurement on the individual signature states \cite{Wallden_14}. The minimum cost $C_{min}$ is the minimum probability that an honest party will detect an error in an individual signature element coming from the forger, and is calculated in the supplemental material \cite{supplementary}. As long as $C_{min}$ is larger than $p_{err}$, which denotes the probability 
{of a mismatch with the sent signature when all parties are honest,}
the signature scheme can be made secure by appropriately choosing other protocol parameters such as the length $L$. Note that $p_{err}$ is determined from experimental data. A final condition for a useful QDS scheme is that it must be robust, i.e. it must succeed with high probability if all parties are honest. 

The exact security definitions can vary and depend on whether one party is more likely to be dishonest than the others. As detailed in the supplemental material~\cite{supplementary}, we 
set protocol parameters so that the repudiation probability, the forging probability and the failure probability are all approximately equal. In this way, the 
probability that the scheme will fail in any one of these ways is bounded by
\begin{equation}
P(\mbox{failure})\le 2\exp\left(-\frac{g^2}{16}L\right), \label{Pfailure}
\end{equation}
where $g=C_{min}-p_{err}$ is the advantage that the legitimate sender Alice has over a forger for a single position of the signature sequence \cite{Donaldson_15,supplementary}. Since the failure probability decays exponentially with the signature length $L$, the scheme is secure, and any required security level can be achieved with sufficiently large $L$. The figure of merit we use to characterise the quality of our QDS schemes is the length $2L$ required to sign a one-bit message with a failure probability of 0.01$\%$.


\begin{figure}[tb]
\includegraphics[width=8cm]{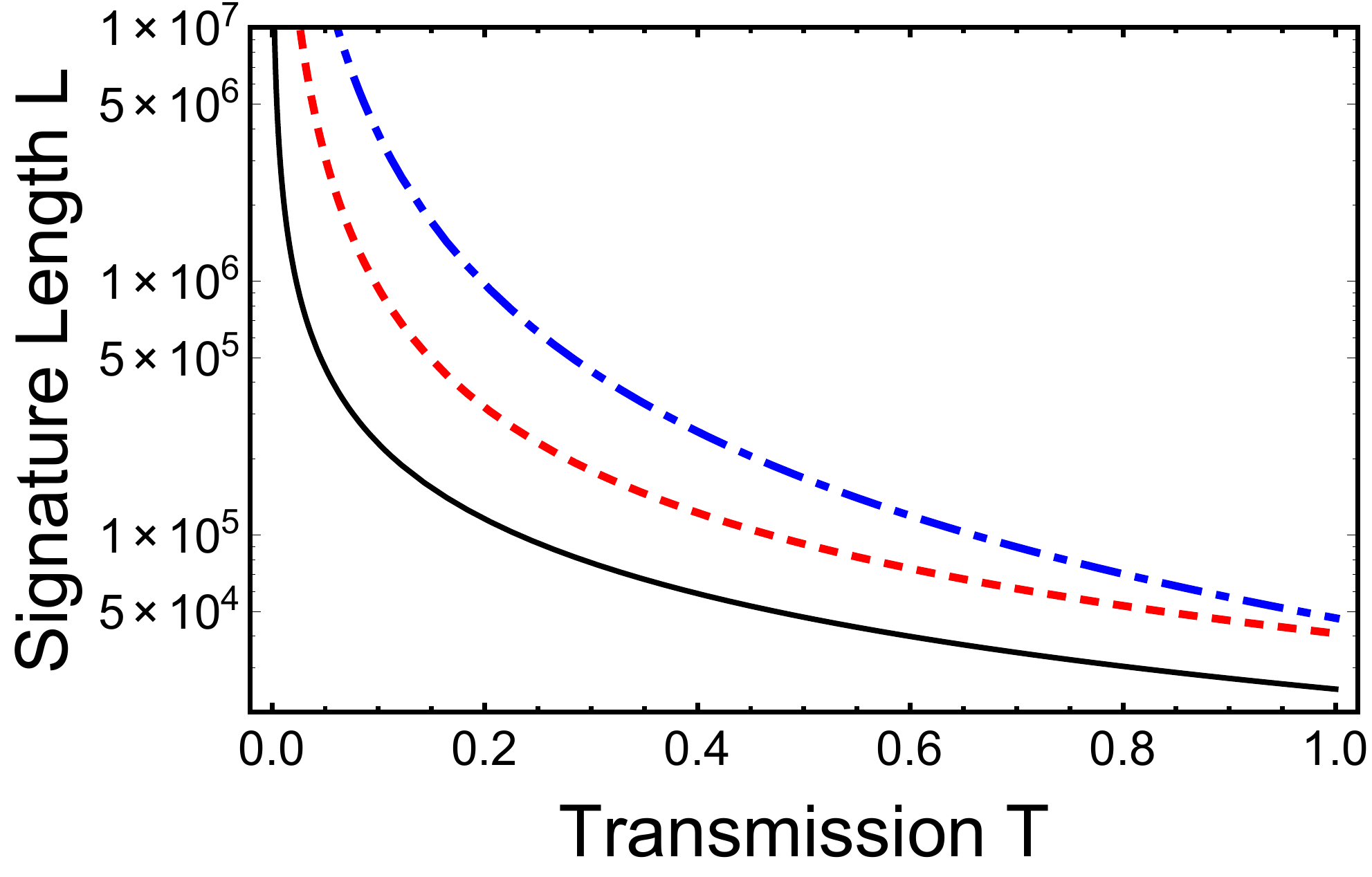}
\caption{Black (solid) curve: Signature length for an ideal ambiguous measurement scheme. Red (dotted) curve: Signature length for an ambiguous measurement scheme with realistic imperfections. Blue (dot-dashed) curve: Signature length for an ideal unambiguous measurement scheme.}\label{fig3}
\end{figure}

{To show the robustness of the protocol}, the experiment was carried out {over a real free-space urban link 
\cite{Peuntinger_14,Heim_14}}.  {The signal states $\lvert \pm \alpha\rangle$,   $\lvert \pm i\alpha\rangle$}  were then repeatedly transmitted, polarization multiplexed with the local oscillator, which is needed for later detection, through a free-space channel between the buildings of the Max Planck Institute and the University of Erlangen-N\"{u}rnberg \cite{Peuntinger_14,Heim_14,Korolkova_02}. The length of the channel is approximately 1.6~km. 
The channel transmission fluctuated between 50\,\% and 85\,\% due to {beam wandering and} scintillation.  At the receiver the signal was split on a balanced beam splitter to measure both the $\hat{x}$ and $\hat{p}$ quadratures. Simultaneously, the transmission was recorded for each state {(for more details  see~\cite{supplementary})}.
The experiment was implemented for three different signal amplitudes, $\alpha=0.48$, $\alpha=0.93$, and $\alpha=1.63$, and we attribute the first (second) half of the measurement time to Bob (Charlie). {To remedy the channel fading}, Bob's (Charlie's) measurement data is then sorted into 32 sub-channels according to the measured transmission \cite{Peuntinger_14,Heim_14}. Depending on the sign of the quadrature measurement values, for each signal state, two of the possible sent states were eliminated.

For each set of data, the sequence of eliminated states was used to produce a cost matrix \cite{Wallden_14} that gives the probability that each state was eliminated for a particular signal state. For each cost matrix, {we calculate the 
minimum difference between an off-diagonal element of the cost matrix  (probability of eliminating a  state {that was not sent}) and the diagonal element of that row (probability of eliminating the sent state)}. 
This difference was multiplied by the appropriate $p_{min}$ to obtain the parameter $g$ from \eqref{Pfailure} for that cost matrix. The minimum probability that a forger will incorrectly identify the state is $p_{min}$ {(see~\cite{supplementary}).} For each $g$, the signature length $2L$ to sign a one-bit message with a failure probability of 0.01$\%$ was calculated. In Fig. \ref{fig1}, the length $L$ is plotted against transmission $T$ {with} $T$+$R$=1 for  
$\alpha=0.48$. 

{To {account for} experimental imperfections, a theoretical model was developed, {using} only experimental data, {with} no free parameters (for details see the supplemental material \cite{supplementary}). {The larger errors bars in Fig.~\ref{fig1} are mostly due to the statistical error of the smaller amount of data available at lower transmission.} The experiment has a clock rate of about $2.2$~MHz and the required signature length of about 
10$^5$ is easily manageable {in the sub-channels}; thus this demonstrates a viable QDS scheme.}


The experiment was also carried out at $\alpha=0.93$ and $\alpha=1.63$ {(results given in \cite{supplementary})}. Increasing $\alpha$ improves the cost matrix but also decreases $p_{min}$, which makes the guess of the forger easier. There is a trade-off between these two effects, with the optimal $\alpha$ predicted to be $\alpha\approx0.5$, supported by the experimental results. 

The main purpose of this experiment is as a test of the measurement procedure used. 
{A} calculation of the cost matrix provides all the information relevant for implementing a full scheme. In the experiment, all the quantum steps were carried out; the rest is classical communication and information processing. 
The experiment is also the first to demonstrate a signature scheme in a free-space setting, in contrast to previous 
experiments using optical fibers.

{It is important to compare the performance of this scheme to previous results}. In \cite{Donaldson_15}, a similar scheme is presented, but with unambiguous state elimination rather than the ``continuous-variable ambiguous state elimination" used here. There, the signature length required was about $10^{9}$, for 500~m of optical fiber and a total loss level of 35\%. Comparing this to our results, the signature length was about $7\times 10^4$ with a similar loss level and a 1.6~km free-space channel. In \cite{Donaldson_15}, the experiment ran at a clock rate of 100~MHz, whereas the clock rate of this experiment was 2.2~MHz. Increasing the clock rate into the GHz range is straightforward with available technology.

{Fig. \ref{fig3} shows the dependence of signature lengths from transmission for the two schemes (details of the models given in~\cite{supplementary}). Even including experimental errors, our scheme requires a shorter signature than the ideal result for~\cite{Donaldson_15}.} {That is, the QDS protocol based on ambiguous state elimination has a fundamental advantage over unambiguous state elimination}. This advantage is even more pronounced when experimental inefficiencies are taken into account. Approximately one order of magnitude of the advantage comes purely from the {chosen} measurement, as shown in Fig.~\ref{fig3}. The rest comes from the improved technical performance of {hom}odyne measurements compared to single-photon detectors.

In conclusion, we have presented a QDS scheme that uses {CV} {hom}odyne measurements. We have experimentally demonstrated that the scheme works over a fluctuating free-space channel. In addition, the {signature} rate per quantum state sent is orders of magnitude better than previous work. Interestingly, despite the ambiguity in the measurements, this scheme has a fundamental advantage over corresponding schemes using  unambiguous measurements.



C. C. and N. K. acknowledge the support from the Scottish Universities Physics Alliance (SUPA) and the Engineering and Physical Sciences Research Council (EPSRC). The project was supported within the framework of the International Max Planck Partnership (IMPP) with Scottish Universities. C.P. and B.H. thank their colleagues at the FAU computer science building for hosting the receiver. E. A. acknowledges the support of EPSRC EP/K015338/1.

\end{document}


\title{Supplemental Material for:\linebreak\textit{\textbf Free-space quantum signatures using {heterodyne} measurements}}

\author{Callum Croal}
\affiliation{School of Physics and Astronomy, University of St.
Andrews, North Haugh, St. Andrews, Fife, KY16 9SS, Scotland}
\author{Christian Peuntinger}
\affiliation{Max Planck Institute for the Science of Light,
G\"unther-Scharowsky-Str. 1/Bldg. 24, Erlangen, Germany}
\affiliation{Institute of Optics, Information and Photonics,
University of Erlangen-Nuremberg, Staudtstra{\ss}e 7/B2, Erlangen,
Germany}
\affiliation{Department of Physics, University of Otago, 730 Cumberland Street, Dunedin, New Zealand}
\author{Bettina Heim}
\affiliation{Max Planck Institute for the Science of Light,
G\"unther-Scharowsky-Str. 1/Bldg. 24, Erlangen, Germany}
\affiliation{Institute of Optics, Information and Photonics,
University of Erlangen-Nuremberg, Staudtstra{\ss}e 7/B2, Erlangen,
Germany}
\author{Imran Khan}
\affiliation{Max Planck Institute for the Science of Light,
G\"unther-Scharowsky-Str. 1/Bldg. 24, Erlangen, Germany}
\affiliation{Institute of Optics, Information and Photonics,
University of Erlangen-Nuremberg, Staudtstra{\ss}e 7/B2, Erlangen,
Germany}
\author{Christoph Marquardt}
\affiliation{Max Planck Institute for the Science of Light,
G\"unther-Scharowsky-Str. 1/Bldg. 24, Erlangen, Germany}
\affiliation{Institute of Optics, Information and Photonics,
University of Erlangen-Nuremberg, Staudtstra{\ss}e 7/B2, Erlangen,
Germany}
\author{Gerd Leuchs}
\affiliation{Max Planck Institute for the Science of Light,
G\"unther-Scharowsky-Str. 1/Bldg. 24, Erlangen, Germany}
\affiliation{Institute of Optics, Information and Photonics,
University of Erlangen-Nuremberg, Staudtstra{\ss}e 7/B2, Erlangen,
Germany}
\author{Petros Wallden}
\affiliation{School of Informatics, University of Edinburgh, 10 Crichton Street, Edinburgh, EH8 9AB, United Kingdom}
\author{Erika Andersson}
\affiliation{SUPA, Institute of Photonics and Quantum Sciences, School of Engineering and Physical Sciences, Heriot-Watt Universiity, David Brewster Building, Edinburgh, EH14 4AS, United Kingdom}
\author{Natalia Korolkova}
\affiliation{School of Physics and Astronomy, University of St.
Andrews, North Haugh, St. Andrews, Fife, KY16 9SS, Scotland}

\date{\today}

\begin{abstract}
This supplemental material is split into four sections. The first gives an intuitive picture of the principle of QDS. The second section shows the security analysis 
for the digital signature scheme. The third section gives details of the experiment and how the experimental graphs were calculated. The fourth section describes how the theoretical models predicting the required signature lengths were calculated.

\end{abstract}

\maketitle

\section{Principle of quantum signatures}

Suppose Alice sends a message to Bob, who in turn may pass it on to Charlie. How can Bob tell that the message is from Alice, and has not been tampered with, and how can Charlie tell that Bob did not modify {or by himself generate} the message? {If one is} using a conventional handwritten signature, {then} Alice has previously distributed copies of her signature, and recipients compare the signature on a document with the previously distributed signature sample. Quantum signature schemes also have two stages, a distribution stage where sequences of quantum states are distributed among the participants, and an entirely ``classical'' messaging stage, which can occur much later, where Alice sends signed messages to Bob or Charlie. In the type of scheme we will employ, Alice distributes sequences of non-orthogonal quantum states to the possible recipients Bob and Charlie, as signatures {for the} possible future messages. The classical information which fully describes these sequences can be viewed as Alice's ``private keys''. Only Alice has exact knowledge of these sequences, i.e. of the private keys. Bob and Charlie, or any other malicious party, are only able to obtain partial information about the quantum states in these sequences, no matter what type of quantum measurements they are using.

The simplest case is if Alice wants to later sign a one-bit message, ``0'' or ``1''.  Alice then initially distributes quantum state sequences corresponding to ``0'' and ``1'', that is, she distributes the quantum signatures for ``0'' and ``1'', respectively. In the messaging stage, if Alice wants to communicate ``0'', she sends the message together with the corresponding private key, that is, the classical description of what states the corresponding sequence contained. Let us say that the recipient is Bob; due to the non-orthogonality of the signature states, Bob inevitably {can obtain only} {partial information about the private key,}  
{whether he is honest and performs the measurements {in} the protocol or not}. Therefore he only has ``noisy'' {information} 
about the private key corresponding to {either ``0'' or ``1"}, no matter what measurement procedure he uses. Note that here ``noisy'' refers {both} to errors in identifying the states, and to any noise in transmission line.  {The latter, however, can in principle be avoided, whereas the former cannot be avoided even in principle.} Bob then checks Alice's classical private key (classical information about the sent quantum signature) against the 
measurement results he obtained in the distribution stage, for the corresponding sequence of quantum states. In practice, due to errors, Bob's measurement results will not perfectly match Alice's private key, but Bob accepts the message as coming from Alice if the ``distance'' between his {stored} ``noisy'' {information} 
and the private key is small enough. {If Bob later wants to forward the message, he} forwards the message (``0'', say) and the corresponding private key {(that he received from Alice)} to Charlie, who tests the signature in the same way as Bob. {Importantly, Bob should be able to know whether Charlie is likely to accept the message already when Bob performs his initial check of the signature, without at that point contacting Charlie.}

{Bob may try to cheat, that is, try to make Charlie accept a forged message as genuinely coming from Alice. This he can do either if he  has received no message from Alice, or if he has received some other message from Alice. For example,
if Bob receives the message ``0'' and corresponding private key from Alice}, he may decide to cheat and {try to convince} Charlie that Alice communicated ``1''. To do so, he would have to send the private key corresponding to ``1'', but this is not at his disposal. Instead, {in the signature protocol we are implementing}, all he can do is send a classical sequence based on 
{the measurement that he performed during the distribution stage, which only gives} a ``noisy'' copy of the private key for ``1''. He should choose the sequence that, {as far as Bob knows}, is most likely to {be accepted by Charlie as coming from Alice.} 
However, Bob's {noisy copy {of the private key}} and Charlie's noisy {private key}
for ``1'' are different, because they are obtained from different measurements. Therefore Charlie will detect a greater {``noise distance''} between his {own} noisy {private key}
and {the one} he receives from Bob, than he would expect if the message came from Alice. Therefore Charlie knows something is wrong and rejects the message.

Alice can also try to cheat by sending a message that Bob will accept but Charlie will reject, i.e., she can try to send a non-transferable message. Bob and Charlie guard against this by symmetrizing their noisy {measurement results for} the private key, {in the distribution stage}. This can be done, for example, by randomly swapping half of their ``noisy'' {measurement results with}
 each other. After this swapping, from Alice's point of view, Bob's and Charlie's measurement results follow the same statistics, and therefore it is impossible for her to create a non-transferable message, as long as Charlie uses a {less strict} threshold for accepting the {signed message} than Bob.

\section{Security analysis}
To be considered a useful scheme, quantum digital signatures (QDS) must be secure against both repudiation and forgery. The scheme is secure if the probability that the signature can be repudiated or forged decays exponentially with the length of the key. In addition, the scheme should be robust, which means that if all parties behave as they should, the protocol runs as intended with high probability. The analysis below follows the same methods as in \cite{Donaldson_15}.

{\it Security against repudiation}: For successful repudiation, Charlie must reject a message that Bob has already accepted. Due to the random swapping of measurement results between Bob and Charlie, the measurement statistics they share are symmetrical,  
which provides security against repudiation. 
No matter what cheating strategy Alice adopts, including strategies involving entangled states, this will result in 
Bob and Charlie having the same probability $p$ to observe a mismatch in the messaging stage.  
Alice can adjust $p$, but this is all she can do.

To achieve successful repudiation, Alice can manipulate the states sent to Bob and Charlie to try to cause a disagreement between them. We give Alice full control over the probability of a mismatch between the private key and Bob's (Charlie's) eliminated signature. We call the probability of a mismatch $p_B$ for states first sent to Bob, and $p_C$ for states first sent to Charlie.

For successful repudiation, Bob must accept the message for both {parts of his signature, each of length $L/2$}, and Charlie has to reject the message {for} at least one part of his signature. Since $P(A\cap B)\le \min\{P(A),P(B)\}$ and $P(A\cup B)\le P(A)+P(B)$, we can write 
\begin{equation}\label{prep1}
\begin{split}
p_{rep}&=P((A\cap B)\cap(C\cup D))\\
&\le \min\{\min\{P(A),P(B)\},P(C)+P(D)\},
\end{split}
\end{equation}
where $P(A)$ ($P(B)$) is the probability that Bob will accept the message using the $L/2$ states received from Alice (Charlie), and $P(C)$ ($P(D)$) is the probability that Charlie will reject the message due to the $L/2$ states received from Bob (Alice).

Using Hoeffding's inequalities~\cite{hoeff}, which bound the probability that the empirical mean of $L$ independent random variables deviates from their expected mean, the probabilities $P(A)$ and $P(B)$ that Bob will accept the message, for the length $L/2$ parts of his eliminated signature received from Alice and Charlie respectively, are
\begin{equation}\begin{split}P(A)&\le\exp[-(p_B-s_a)^2L],\\
P(B)&\le\exp[-(p_C-s_a)^2L],\end{split}\end{equation}
where $s_a$ is the authentication threshold.
Similarly, the probabilities $P(C)$ and $P(D)$ that Charlie will reject the message for the length $L/2$ parts of his eliminated signature received from Bob and Alice respectively are
\begin{equation}\begin{split}P(C)&\le\exp[-(s_v-p_B)^2L],\\
P(D)&\le\exp[-(s_v-p_C)^2L],\end{split}\end{equation}
 where $s_v$ is the verification threshold and $s_v>s_a$.

Now we can take $p=\max\{p_B,p_C\}$. In that case $\exp[-(p-s_a)^2L]=\min\{P(A),P(B)\}$. In addition, $2\exp[-(s_v-p)^2L]\ge P(C)+P(D)$. Combining these two equations with Eq. \eqref{prep1}, we get
\begin{equation}
p_{rep}\le\min\{2\exp[-(p-s_a)^2L],2\exp[-(s_v-p)^2L]\},
\end{equation}
where the first term in the {minimum} has been doubled for simplicity, noting that this slightly loosens the tightness of the bound on the repudiation probability.

Alice's optimal choice of $p$ is the one that maximizes the smaller of these two terms, that is, $p=\frac{s_a+s_v}{2}$.
With this choice, her repudiation probability is bounded as
\begin{equation}
p_{rep}\le2\exp\left[-\frac{(s_v-s_a)^2}{4}L\right].
\label{prep}
\end{equation}
This decays exponentially with the length of the signature and thus the scheme is secure against repudiation.

{\it Security against forging}: 
It is easier to forge a message that is claimed to be forwarded, than one that is claimed to come directly from Alice. Bounding the probability for the former also bounds the probability for the latter. Therefore, we will consider the case where Bob attempts to forge a message  which he is forwarding to Charlie, claiming he received it from Alice. Since the protocol is symmetric with respect to the two recipients Bob and Charlie, this also bounds Charlie's probability to forge messages.

To successfully forge, Bob must ensure that he doesn't, in the messaging stage, declare {too many} of the states that Charlie has eliminated, with fewer than $s_vL/2$ errors in each length $L/2$ part of Charlie's eliminated signature. Since Bob can control what he forwards to Charlie in the distribution stage, Bob can completely control the number of mismatches for these positions. If he so wishes, he can cause no mismatches in those positions. 
Therefore it is the measurement results which Charlie did not forward to Bob that Bob has to try to guess. 
The measurement results Charlie received through Bob are used to protect against repudiation, whereas the measurement results Charlie obtained for states directly received from Alice are used to test for forgery by Bob, and vice versa.

Assuming that Bob cannot interfere with the quantum states which Alice sends to Charlie, Bob's best forging strategy will involve measurements on the copies of these states which Bob legitimately received from Alice. Based on this, Bob will make a best guess, when later declaring to Charlie what these states supposedly were. The optimal measurement Bob should make to forge is limited only by what is possible in quantum mechanics, not by any considerations of what measurements are practical to realize, and is not the same measurement as he would make if honestly following the protocol. In general, one should assume that Bob knows which measurement results Charlie will forward, and which ones he will keep to himself, so that Bob can change his measurement strategy accordingly for states in different positions.

The fact that the possible states Alice can send are non-orthogonal provides the basis of the security of the scheme. As in \cite{Collins_14}, the optimal individual measurement Bob can perform is a minimum-cost measurement, minimising Bob's ``cost" associated with mismatches. Since the states sent by Alice are uncorrelated with each other, collective forging strategies, where measurements on successive signature states can depend on the results obtained in previous measurements, provide no advantage over individual forging strategies, where Bob simply repeats the same optimal measurement for each signature state~\cite{Collins_14}. The most general type of forging attack are coherent forging attacks, where Bob can measure any number of signature states in an entangled basis. While intuitively the protocol should remain secure also against coherent forging, this analysis is not in general straightforward. We therefore leave discussion of coherent forging attacks for future work, noting that it has  been shown that for BB84 signature states, coherent attacks provide no advantage \cite{Wallden_14}. 

To prove security against individual and collective forging, we need to bound Bob's minimum cost for a measurement on an individual signature state, which in this case is identical to Bob's probability to cause a mismatch for a single signature element. This is done following the method in the supplemental material of \cite{Donaldson_15}, resulting in a lower bound on the minimum cost $C_{min}$, depending on the cost matrix, which is determined from the experimental data, and
 $p_{min}$, which is the minimum probability for Bob to incorrectly identify a state received from Alice. $p_{min}$ depends on the amplitude of the initial coherent states, and can be shown to be \cite{Collins_14}
\begin{equation}
p_{min}=1-\frac{1}{16}|\sum_{i=1}^4\sqrt{\lambda_i}|^2,\label{pmin}
\end{equation}
where
$\lambda_{1,2}=2\exp(-\alpha^2)[\cosh(\alpha^2)\pm\cos(\alpha^2)]$  and $\lambda_{3,4}=2\exp(-\alpha^2)[\sinh(\alpha^2)\pm\sin(\alpha)^2].$ 
Here, we are assuming that the forger Bob has access to the states Alice sends before any losses or imperfections have acted on them. This is not true for an honest Charlie, whose measurements on the states is subject to loss and imperfections. An example of a calculation of a bound for the minimum cost for an experimental cost matrix is given in Section 2, and a calculation for a theoretical cost matrix is given in Section 3.

The probability of a successful forgery is the probability that Charlie measures fewer than $s_v L/2$ errors in the results for the $L/2$ states received directly from Alice during forgery by Bob. Using Hoeffding's inequalities
, the probability of a successful forgery is therefore
\begin{equation}
p_{forg}\le\exp\left[-(C_{min}-s_v)^2L\right].\label{pforg}
\end{equation}
This probability decays exponentially with respect to signature length as long as $C_{min}>s_v$.

{\it Robustness}: A QDS scheme is only useful if it only fails with small probability. If all parties are honest, then Bob should accept the message as being genuine, except with small probability. The message is rejected if Bob detects more than $s_aL/2$ errors in either of the length $L/2$ parts of his eliminated signature, which using Hoeffding's  inequalities 
occurs with probability
\begin{equation}
p_{fail}\le2\exp\left[-(s_a-p_{err})^2L\right],\label{pfail}
\end{equation}
where $p_{err}$ is the probability that an honest recipient, following the protocol, will eliminate the state actually sent by Alice. If, as is normally the case, $p_{err}$ for the states sent to Charlie is different to that for those sent to Bob, then $p_{err}$ should be taken as the maximum of those probabilities.
Since Charlie's rejection threshold is less strict than Bob's, Charlie's rejection probability is smaller than Bob's. 
For the protocol to be robust, we thus have to choose $s_v>s_a>p_{err}$.

Taking everything together, the protocol can be made secure and robust as long as an honest Charlie is able to distinguish a ``fake" declaration by Bob from a declaration made by Alice, in terms of the average number of mismatches Charlie sees. This occurs when Bob's optimum probability to cause a mismatch, $C_{min}$, is greater than the probability $p_{err}$ that Alice's true declaration will cause a mismatch.  
As long as $C_{min} > p_{err}$ holds, the thresholds $s_v$, $s_a$ and the signature length $L$ can be chosen so that the scheme is as secure as desired against forging for all displacement amplitudes.

If we assume that all parties are equally likely to be dishonest, then we can define the level of security by setting the terms in the exponentials of  Eqs. \eqref{prep}, \eqref{pforg} and \eqref{pfail} to be equal to each other. This is achieved when $s_a=p_{err}+(C_{min}-p_{err})/4$, and $s_v=p_{err}+3(C_{min}-p_{err})/4$. This gives an upper bound for the total probability for the scheme to fail in any one of these ways of
\begin{equation}
P(\mbox{failure})\le2\exp\left(-\frac{g^2}{16}L\right),
\label{failureprob}
\end{equation}
where $g=C_{min}-p_{err}$ can be determined 
from experimental results. 
The figure of merit we use to characterize the quality of a QDS scheme is the length $2L$ required to sign a one-bit message for a particular security level. In this work, to facilitate comparison with earlier realizations~\cite{Collins_14, Donaldson_15}, the security level we choose is that the probability of failure is $\le$ 0.01$\%$.

\section{Experimental Details}

{For the signature state sequences} we use four 
coherent states  $\lvert \alpha\rangle$, $\lvert i\alpha\rangle$, $\lvert -\alpha\rangle$, $\lvert -i\alpha\rangle$,  which are symmetrically distributed in quadrature phase space.
In \cite{Heim_14},
we used these same states to distribute and quantify effective entanglement between Alice and Bob. 
At the receiver, we measure both the $\hat{x}$ and the $\hat{p}$ quadrature using a heterodyne measurement. The signal is split at a balanced beam splitter, and homodyne detectors are used at both outputs. In particular, in each homodyne measurement, we mix a strong local oscillator with the signal on a balanced beam splitter, and measure the resulting difference signal of two PIN photodiodes, built into a homemade detector. To achieve a high detection efficiency at the receiver we send the local oscillator (LO) together with the signal states, polarization multiplexed, through the 1.6\,km free space channel. This can be described using Stokes operators \cite{Korolkova_02}. 

At Alice's end, we use a grating-stabilized diode laser at 809~nm wavelength (Toptica DL 100). The output of this laser is spatially mode-cleaned by a single-mode fiber, and a small part of the output power is used in a balanced self-homodyning setup to monitor the shot noise-limited operation of the laser. The remaining part is used to prepare the actual signals and also the local oscillator. The polarization is cleaned by a polarizing beam splitter (PBS) and then adjusted to be circularly polarized ($\langle\hat{S}_1\rangle=\langle\hat{S}_2\rangle=0$) using a quarter-wave plate (QWP). Then we use two sequences of half-wave plates (HWP) and electro-optical modulators (EOM, Thorlabs, EO-AM-NR-C1, 600-900 nm, bandwidth 100 MHz) to produce the four signal states in the $S_1$-$S_2$-plane. In terms of Stokes operators this leads to a bright $+S_3$-polarized local oscillator (which is essentially not affected by the signal modulation) and the signal states $\lvert \alpha\rangle$, $\lvert i\alpha\rangle$, $\lvert -\alpha\rangle$, $\lvert -i\alpha\rangle$, which are orthogonally (i.e.\ $-S_3$-) polarized. Therefore the EOMs are driven by two individual arbitrary waveform generators (Agilent 33250A) that are synchronized with each other. They are used to produce Gaussian-shaped modulation voltages with peak voltage in the mV range, leading to signal amplitudes in the range of a few shot-noise units. After each Gaussian-shaped pulse, the output voltage is set to zero for the same time period. This is used as the vacuum reference for the signal states. The repetition rate of the produced signal states is 3.05~MHz. After 263 signal pulses, we increase the peak voltage of one pulse to produce a trigger signal for synchronization between Alice and Bob (Charlie). To avoid any influence of the bright trigger pulse onto the quantum signals, we disregard the 92 following signal pulses. This leads to an effective sending rate of 2.22~MHz. At the sender, the signal preparation is either verified, or the beam is expanded to a beam width of approximately 4~cm and sent through an optical window followed by the free-space channel to Bob (Charlie). The signal measurement at Alice, used to adjust and confirm the signal preparation, uses a balanced beam splitter to split the signals in two equal parts. They are mixed with the polarization-multiplexed LO on a PBS. The phase of the LO can in this case be adjusted using a HWP, while a QWP is used to compensate for static polarization offsets. The outputs of the PBS are detected with two PIN photodiodes and the difference signal of these are amplified in a homemade detector. By this we are able to simultaneously measure the $S_1$ ($\hat{x}$ quadrature) observable and the $S_2$ ($\hat{p}$ quadrature) observable. The overall detection efficiency, including optical losses and the diodes' quantum efficiencies, is $0.84\pm0.02$. The electronic signal is high-pass filtered  (Minicircuits BLK-89-S+, 100 kHz) and analogue-to-digital-converted with an oscilloscope at a sampling rate of 250 Msamples/s. Thus each signal pulse is 41 samples long, followed by 41 samples of vacuum. The linearity of the detection system was confirmed by an attenuation measurement without signal modulation.

At Bob (Charlie) we use a telescope with a receiving aperture of 150~mm to catch as much as possible of the incoming beam and reduce its beam width for further processing. First we split 5\% from the beam with an unbalanced beam splitter, to record the channel transmission, which was between 50\% and 85\% during our measurements. The remaining received signal is measured in exactly the same manner as at Alice's site. Here the overall detection efficiency including optical losses and the diodes' quantum efficiencies is $0.83\pm0.02$. The experiment was implemented with three different signal peak voltages leading to the average signal sizes $\alpha=0.48$, $\alpha=0.93$, and $\alpha=1.63$. The $S_1$ signals are slightly reduced compared to the $S_2$ signals, as we use the same modulation voltages but produce the $S_1$ signals first. Thus they are attenuated by the second EOM which has a transmittance of 95\%. We attribute the first (second) half of the overall measurement time to Bob (Charlie). As already mentioned in the main article Bob's (Charlie's) measurement data is then sorted in 32 sub-channels, according to the measured transmission. Depending on the sign of the quadrature measurement values, for each signal state two of the possibly sent states were eliminated. For example, in the case of a positive $S_1$  ($S_2$) measurement value, $\lvert-\alpha\rangle$ ($\lvert-i\alpha\rangle$) is eliminated.

For each $\alpha$ and transmission bin, the knowledge of the sent state was combined with the eliminated states to produce a cost matrix that gives the probability that each state was eliminated for each sent state. The rows of the matrix correspond to the states sent by Alice, in the order $\lvert \alpha\rangle$, $\lvert i\alpha\rangle$, $\lvert -\alpha\rangle$, $\lvert -i\alpha\rangle$. The columns correspond to the states eliminated by Bob in the same order. The diagonal elements therefore give the probability that the sent state is eliminated. An example of the measured cost matrix is shown below, with errors. {The errors are calculated by dividing the available dataset in 10 equal sized parts and calculating the respective cost
matrix and their standard deviation. Thus the errors give an upper bound for the statistical error and possibly drifting systematic errors.} This matrix is Bob's data for $\alpha=0.48$ at a transmission level of $T=0.600$ ($T$+$R$=1)  and is given by
\[C=\left(\begin{array}{cccc}
0.3767&0.5028&0.6233&0.4972\\
0.4929&0.3682&0.5071&0.6318\\
0.5979&0.496&0.4021&0.504\\
0.4957&0.6204&0.5043&0.3796\\
\end{array}\right)\]
\begin{equation}\pm\left(\begin{array}{cccc}
0.015&0.019&0.015&0.019\\
0.008&0.013&0.008&0.013\\
0.013&0.019&0.013&0.019\\
0.014&0.020&0.014&0.020\\
\end{array}\right).\label{ECM}\end{equation}
The relevant cost matrix can be used to bound the minimum cost of a minimum-cost measurement performed by a forger, by following the method in the supplemental material of \cite{Collins_14}. 

{To find an analytical bound on the minimum cost, we manipulate the cost matrix in Eq. \eqref{ECM} to the form of an error-type cost matrix. We do this because the minimum cost of an error-type cost matrix is proportional to $p_{min}$, the minimum probability to incorrectly identify the state, with the proportionality given by the off-diagonal elements of the cost matrix. An error-type cost matrix has zeros on the diagonals of the cost matrix, and all the off-diagonal terms are equal. It is called error-type because a correct declaration has zero cost, and an incorrect declaration always has the same cost.}

{To get to this form, we use two properties of cost matrices. First, subtracting a constant-row matrix from a cost matrix reduces the cost by a constant, while leaving the minimum-cost measurement unchanged. Second, the cost of a cost matrix $C_{i,j}$ is 
{bounded from below} by the cost of a cost matrix $C_{i,j}^l$ that is strictly smaller than it $C_{i,j}^l\le C_{i,j}$.}

\begin{figure}[tb]
\includegraphics[width=8cm]{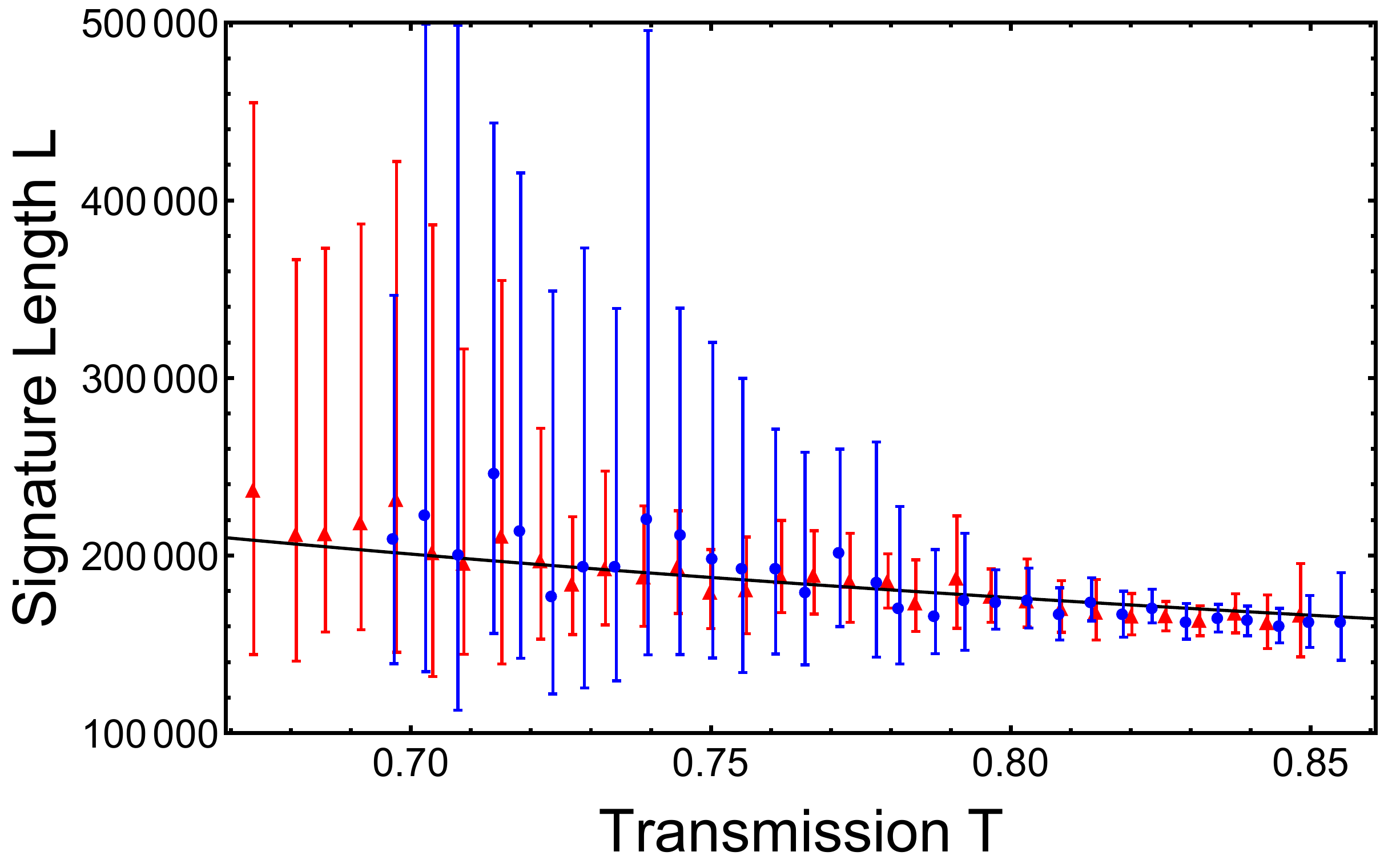}
\caption{Signature length for $\alpha=0.93$. Blue curve: theoretical model. Blue dots/bars: results from the data attributed to Bob. Red dots/bars: results from the data attributed to Charlie. The error bars calculated are statistical.}\label{fig2}
\end{figure}

\begin{figure}[tb]
\includegraphics[width=8cm]{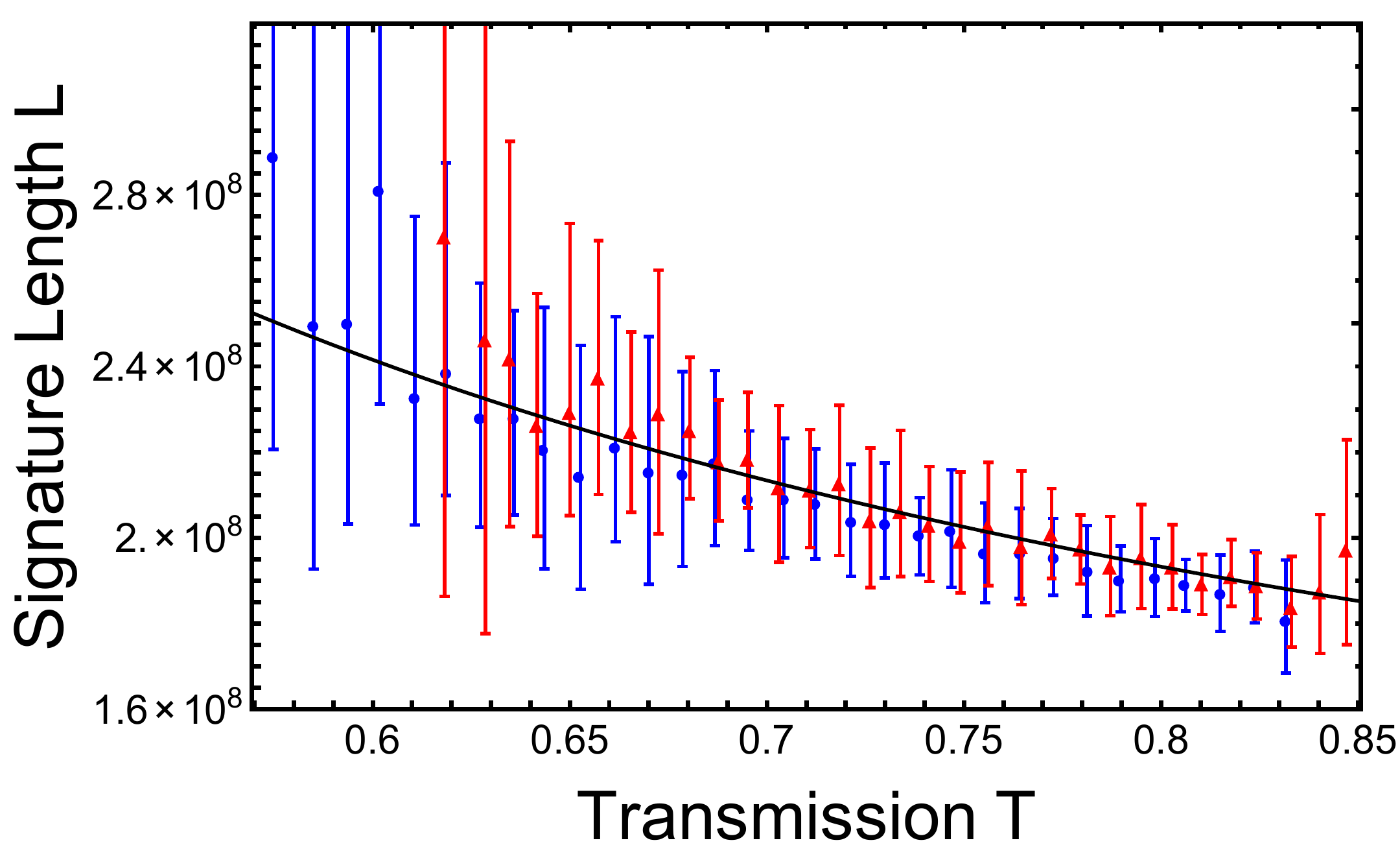}
\caption{Signature length for $\alpha=1.63$. Blue curve: theoretical model. Blue dots/bars: results from the data attributed to Bob. Red dots/bars: results from the data attributed to Charlie. The error bars calculated are statistical.}\label{fig3}
\end{figure}

{We define $C_{i,j}^h=C_{i,i}$, a constant-row matrix for which the elements in each row are equal to the diagonal elements of the matrix $C_{i,j}$. We then define $C_{i,j}'=C_{i,j}-C_{i,j}^h$, which has the same minimum-cost measurement as $C_{i,j}$, but with the minimum cost reduced by $C^{h}=1/4\sum_iC_{i,i}$. Finally we define the cost matrix $C_{i,j}^l$ that is strictly smaller than $C_{i,j}'$ for all $i,j$ such that $C_{i,j}^l=\min_{i\ne j}C_{i,j}'$ for all $i\ne j$, and with zeros on the diagonal. This final cost matrix $C_{i,j}^l$ is of error-type, for which the minimum cost $C_{min}^l$ is proportional to the minimum error probability $p_{min}$. Using this argument we can lower bound the minimum cost of the cost matrix \eqref{ECM} as}
\begin{equation}
C_{min}\ge C^h+C_{min}^l.
\end{equation}
Starting from \eqref{ECM}, the subsequent cost matrices are
\begin{equation}
C^h=\left(\begin{array}{cccc}
0.3767&0.3767&0.3767&0.3767\\
0.3682&0.3682&0.3682&0.3682\\
0.4021&0.4021&0.4021&0.4021\\
0.3796&0.3796&0.3796&0.3796\\
\end{array}\right),\label{constantrow}\end{equation}
\begin{equation}
C'=\left(\begin{array}{cccc}
0&0.1261&0.2466&0.1205\\
0.1247&0&0.1389&0.2636\\
0.1958&0.0939&0&0.1019\\
0.1161&0.2408&0.1247&0\\
\end{array}\right),\end{equation}
\begin{equation}
C^l=\left(\begin{array}{cccc}
0&0.0939&0.0939&0.0939\\
0.0939&0&0.0939&0.0939\\
0.0939&0.0939&0&0.0939\\
0.0939&0.0939&0.0939&0\\
\end{array}\right).\label{Cl}\end{equation}
From \eqref{constantrow}, $C^h=0.3817$. This is the cost for an honest scenario; it is the probability that Charlie will eliminate a state that Alice sent if all parties are honest. From \eqref{Cl}, the minimum difference between the probability of eliminating the sent state, and the probability of eliminating another state is 0.0939. This difference therefore gives the advantage of declaring the sent state at the messaging stage.
The minimum cost for matrix \eqref{Cl} is the product of that advantage and the minimum probability to incorrectly identify a state $p_{min}$. For this state $\alpha=0.48$, so from \eqref{pmin}, $p_{min}=0.4373$. The minimum cost of the matrix $C_{i,j}$ is finally
\begin{equation}
C_{min}=0.3817+0.0939\times0.4373=0.42276,
\end{equation}
and the parameter $g$ used to calculate the signature length is
\begin{equation}
g=C_{min}-C^h=0.04106.
\end{equation}
This corresponds to a required signature length of $L=94000$ for a security level of 0.01\%.
In Figs. \ref{fig2} and \ref{fig3}, we plot similarly calculated signature lengths as a function of the channel transmission, for $\alpha=0.93$ and $\alpha=1.63$. The graph for $\alpha=0.48$ is already included in the main paper.

In all experimental graphs, errors in the signature length were calculated using the statistical errors of the elements in the cost matrices. The errors in the length were calculated by first adding the errors of the diagonal elements, and subtracting the errors of the off-diagonal elements. This gives a new cost matrix $C'$ from which a new parameter $g'$ can be calculated as above, with $g'<g$. This new $g'$ is then used to calculate a new length $L'>L$ that is the worst-case scenario for the required signature length. The length $L'$ is taken as the top of the error bar in the experimental graph.

Second, the error bars in the diagonal elements are subtracted, and the errors in the off-diagonal elements are added to give a new cost matrix $C''$ that has a new parameter $g''>g$. This new $g''$ is then used to calculate a new length $L''<L$ that gives a best-case scenario for the required signature length. The length $L''$ is used for the bottom of the error bar in the experimental graph.

Note, that to ensure the required security when running a full signature protocol, the longest length $L'$ should be used for the signature length, as this is the worst-case scenario. This means it is important to minimize the errors in the cost matrix by taking a large number of measurements to calculate the cost matrix. In this experiment, insufficient data was available at some transmission levels, which led to the large error bars seen.

\section{Theoretical Models}

In Fig. 3 of the main paper, the required signature length required with respect to transmission $T$ for three different theoretical models is shown. Here we describe how those curves were calculated.

The black (lower) curve shows the case where heterodyne detection is used by the honest recipients and there are no experimental imperfections.  In this case, the ideal cost matrix is
\begin{equation}
C=\left(\begin{array}{cccc}
p_{err}&1/2&1-p_{err}&1/2\\
1/2&p_{err}&1/2&1-p_{err}\\
1-p_{err}&1/2&p_{err}&1/2\\
1/2&1-p_{err}&1/2&p_{err}\\
\end{array}\right),\label{theoryCM}\end{equation}
where $p_{err}=\frac{1}{2}\mbox{erfc}\left(\sqrt{\frac{T}{2}}\alpha\right).$ From this cost matrix, the minimum cost is bounded as described for the experimental cost matrix in the previous section. In this way, the minimum cost is found to be $C_{min}=p_{err}+p_{min}(\frac{1}{2}-p_{err})$, and the parameter $g$ is thus $g=p_{min}(\frac{1}{2}-p_{err})$. Note that the $\alpha$ used to calculate $p_{min}$ is the unattenuated $\alpha$ prepared by Alice. A higher $g$ gives a shorter signature length and therefore the optimal $\alpha$ is the one that gives the highest $g$. In this case $g$ is maximal when $\alpha\approx0.5$. The black curve calculated in Fig. 3 of the main paper is plotted by fixing $\alpha=0.5$ and calculating $L$ from the resulting $g$.

The red (middle) curve in Fig. 3 of the main paper shows the case where heterodyne detection is used by the honest recipients and some experimental imperfections are taken into account. The experimental imperfections considered are imperfect detection efficiency, additional variance introduced by the EOM that displaces the coherent states, and electronic noise that increases the variance at the measurement stage. When these imperfections are taken into account, the cost matrix is the same as in Eq. \eqref{theoryCM}, but with a new $p_{err}$,
\begin{equation}p_{err}=\frac{1}{2}\mbox{erfc}\left(\frac{\frac{1}{2}\eta T\alpha}{\sqrt{\frac{1}{2}\eta T\epsilon+elect}}\right),\end{equation}
where $\eta$ is the detection efficiency, $\epsilon$ is the additional variance that comes in from the state prepartation and $elect$ is the electronic noise that increases the variance of the states. In all experiments, $\eta=0.856$ and $\epsilon=1.01$, and $elect$ varies between 0.04 and 0.08. The value of $elect$ is determined from the measured variances of the states. The theoretical model also takes into account the fact that the modulation of the Stokes operators $\hat{S}_1$ and $\hat{S}_2$ had a slightly different amplitude. The lower amplitude of $\hat{S}_1$ was used to calculate the guaranteed advantage from the cost matrix, and the higher amplitude of $\hat{S}_2$ was used to calculate $p_{min}$. The encoding always has some phase imperfections, however, since this only has a small effect on the signature length, it is not included in the model for simplicity. The signature length is calculated from the cost matrix in the same way as for the black curve, and the result is plotted in Fig. 3 of the main paper. This model is the same one that was used to plot the theoretical curve for Fig. 2 of the main paper and Figs. \ref{fig2} and \ref{fig3} of the supplementary material.

The blue (upper) curve shows the case where single-photon detection is used for unambiguous state elimination as in \cite{Collins_14}, and there are no experimental imperfections. This represents the optimum length achievable for these states using unambiguous state elimination. In this case the ideal cost matrix is
\begin{equation}
C=\left(\begin{array}{cccc}
0&q&p&q\\
q&0&q&p\\
p&q&0&q\\
q&p&q&0\\
\end{array}\right),\label{theoryCM2}\end{equation}
where 
$$p=1-\exp(-|\sqrt{T}\alpha|^2), ~~q=1-\exp(-|\sqrt{T}\alpha|^2/2).$$ From this, the minimum cost is bounded as before to be $C_{min}=p_{min}q$. Since the diagonal elements are 0, $C_{min}=g$, and $g$ is used to calculate the required signature length. Again, a higher $g$ gives a shorter signature length and therefore the optimal $\alpha$ is the one that gives the highest $g$. The blue curve is plotted by using the optimal $\alpha$ at each level of transmission, which in this case is $\alpha\approx0.7$.

The black and blue curves can be compared to study which measurement scheme is most efficient for this set of states. They both assume ideal experimental conditions and so remove any technical considerations. Since the black curve is always lower than the blue curve, this show that heterodyne detection has a fundamental advantage over single photon detection for this protocol. In fact, even when taking into account realistic experimental imperfections, the scheme based on heterodyne detection performs better than the one based on single photon detection could ever do.